\documentclass[reprint,
 amsmath,amssymb,
 aps,prd,
 superscriptaddress,
]{revtex4-1}

\usepackage{graphicx}% Include figure files
\usepackage{dcolumn}% Align table columns on decimal point
\usepackage{lineno}
\usepackage{bm}% bold math
\usepackage{comment}
\usepackage{xspace}
\usepackage[dvipsnames]{xcolor}
\usepackage{units}
\usepackage{hyperref}
\usepackage{url}
\usepackage{float}
\usepackage[normalem]{ulem}
\newcommand{\ANL}{Argonne National Laboratory (ANL), Lemont, IL, 60439, USA}
\newcommand{\Bern}{Universit{\"a}t Bern, Bern CH-3012, Switzerland}
\newcommand{\BNL}{Brookhaven National Laboratory (BNL), Upton, NY, 11973, USA}
\newcommand{\UCSB}{University of California, Santa Barbara, CA, 93106, USA}
\newcommand{\Cambridge}{University of Cambridge, Cambridge CB3 0HE, United Kingdom}
\newcommand{\CIEMAT}{Centro de Investigaciones Energ\'{e}ticas, Medioambientales y Tecnol\'{o}gicas (CIEMAT), Madrid E-28040, Spain}
\newcommand{\Chicago}{University of Chicago, Chicago, IL, 60637, USA}
\newcommand{\Cincinnati}{University of Cincinnati, Cincinnati, OH, 45221, USA}
\newcommand{\CSU}{Colorado State University, Fort Collins, CO, 80523, USA}
\newcommand{\Columbia}{Columbia University, New York, NY, 10027, USA}
\newcommand{\Edinburgh}{University of Edinburgh, Edinburgh EH9 3FD, United Kingdom}
\newcommand{\FNAL}{Fermi National Accelerator Laboratory (FNAL), Batavia, IL 60510, USA}
\newcommand{\Granada}{Universidad de Granada, Granada E-18071, Spain}
\newcommand{\Harvard}{Harvard University, Cambridge, MA 02138, USA}
\newcommand{\IIT}{Illinois Institute of Technology (IIT), Chicago, IL 60616, USA}
\newcommand{\KSU}{Kansas State University (KSU), Manhattan, KS, 66506, USA}
\newcommand{\Lancaster}{Lancaster University, Lancaster LA1 4YW, United Kingdom}
\newcommand{\LANL}{Los Alamos National Laboratory (LANL), Los Alamos, NM, 87545, USA}
\newcommand{\Louisiana}{Louisiana State University, Baton Rouge, LA, 70803, USA}
\newcommand{\Manchester}{The University of Manchester, Manchester M13 9PL, United Kingdom}
\newcommand{\MIT}{Massachusetts Institute of Technology (MIT), Cambridge, MA, 02139, USA}
\newcommand{\Michigan}{University of Michigan, Ann Arbor, MI, 48109, USA}
\newcommand{\Minnesota}{University of Minnesota, Minneapolis, MN, 55455, USA}
\newcommand{\NMSU}{New Mexico State University (NMSU), Las Cruces, NM, 88003, USA}
\newcommand{\Oxford}{University of Oxford, Oxford OX1 3RH, United Kingdom}
\newcommand{\Pitt}{University of Pittsburgh, Pittsburgh, PA, 15260, USA}
\newcommand{\Rutgers}{Rutgers University, Piscataway, NJ, 08854, USA}
\newcommand{\SLAC}{SLAC National Accelerator Laboratory, Menlo Park, CA, 94025, USA}
\newcommand{\SDSMT}{South Dakota School of Mines and Technology (SDSMT), Rapid City, SD, 57701, USA}
\newcommand{\Maine}{University of Southern Maine, Portland, ME, 04104, USA}
\newcommand{\Syracuse}{Syracuse University, Syracuse, NY, 13244, USA}
\newcommand{\TelAviv}{Tel Aviv University, Tel Aviv, Israel, 69978}
\newcommand{\Tennessee}{University of Tennessee, Knoxville, TN, 37996, USA}
\newcommand{\UTA}{University of Texas, Arlington, TX, 76019, USA}
\newcommand{\Tufts}{Tufts University, Medford, MA, 02155, USA}
\newcommand{\UCL}{University College London, London WC1E 6BT, United Kingdom}
\newcommand{\VTech}{Center for Neutrino Physics, Virginia Tech, Blacksburg, VA, 24061, USA}
\newcommand{\Warwick}{University of Warwick, Coventry CV4 7AL, United Kingdom}
\newcommand{\Yale}{Wright Laboratory, Department of Physics, Yale University, New Haven, CT, 06520, USA}
\begin{document}

%\linenumbers

\preprint{APS/123-QED}

\title{First measurement of $\eta$ production in neutrino interactions on argon with MicroBooNE}% Force line breaks with \\
%\thanks{A footnote to the article title}%

% List of institutions in command form:

%%\newcommand{\listerThanks}{Now at University of Wisconsin, Madison}

% So that institutions appear in alphabetical order:
\affiliation{\ANL}
\affiliation{\Bern}
\affiliation{\BNL}
\affiliation{\UCSB}
\affiliation{\Cambridge}
\affiliation{\CIEMAT}
\affiliation{\Chicago}
\affiliation{\Cincinnati}
\affiliation{\CSU}
\affiliation{\Columbia}
\affiliation{\Edinburgh}
\affiliation{\FNAL}
\affiliation{\Granada}
\affiliation{\Harvard}
\affiliation{\IIT}
\affiliation{\KSU}
\affiliation{\Lancaster}
\affiliation{\LANL}
\affiliation{\Louisiana}
\affiliation{\Manchester}
\affiliation{\MIT}
\affiliation{\Michigan}
\affiliation{\Minnesota}
\affiliation{\NMSU}
\affiliation{\Oxford}
\affiliation{\Pitt}
\affiliation{\Rutgers}
\affiliation{\SLAC}
\affiliation{\SDSMT}
\affiliation{\Maine}
\affiliation{\Syracuse}
\affiliation{\TelAviv}
\affiliation{\Tennessee}
\affiliation{\UTA}
\affiliation{\Tufts}
\affiliation{\UCL}
\affiliation{\VTech}
\affiliation{\Warwick}
\affiliation{\Yale}

% Authors in alphabetical order
\author{P.~Abratenko} \affiliation{\Tufts}
\author{O.~Alterkait} \affiliation{\Tufts}
\author{D.~Andrade~Aldana} \affiliation{\IIT}
\author{J.~Anthony} \affiliation{\Cambridge}
\author{L.~Arellano} \affiliation{\Manchester}
\author{J.~Asaadi} \affiliation{\UTA}
\author{A.~Ashkenazi}\affiliation{\TelAviv}
\author{S.~Balasubramanian}\affiliation{\FNAL}
\author{B.~Baller} \affiliation{\FNAL}
\author{G.~Barr} \affiliation{\Oxford}
% ...

\collaboration{The MicroBooNE Collaboration}
\thanks{microboone\_info@fnal.gov}\noaffiliation
%\email[]{microboone\_info@fnal.gov}

\begin{abstract}
We present a measurement of $\eta$ production from neutrino interactions on argon with the MicroBooNE detector. The modeling of resonant neutrino interactions on argon is a critical aspect of the neutrino oscillation physics program being carried out by the DUNE and Short Baseline Neutrino programs. $\eta$ production in neutrino interactions provides a powerful new probe of resonant interactions, complementary to pion channels, and is particularly suited to the study of higher-order resonances beyond the $\Delta(1232)$. We measure a flux-integrated cross section for neutrino-induced $\eta$ production on argon of $3.22 \pm 0.84 \; \textrm{(stat.)} \pm 0.86 \; \textrm{(syst.)}$ $10^{-41}{\textrm{cm}}^{2}$/nucleon.  By demonstrating the successful reconstruction of the two photons resulting from $\eta$ production, this analysis enables a novel calibration technique for electromagnetic showers in GeV accelerator neutrino experiments.
\end{abstract}

%\keywords{Suggested keywords}%Use showkeys class option if keyword
%display desired
\maketitle

%\tableofcontents

%The study of neutrinos, a particle in the Standard Model (SM) of particle physics, is a key component of the particle physics community's aim to understand the fundamental laws of nature.
Neutrino oscillation physics experiments have embarked on an expansive program aimed at performing precision measurements of neutrino oscillation parameters including measurements of the Charge-Parity violating phase in the lepton sector, $\delta_{\textrm{CP}}$. These experiments additionally provide a unique environment to search for new physics through possible rare processes occurring along the beamline. This research program is in part enabled by the accelerator-based neutrino oscillation program which leverages GeV-scale neutrino beams and liquid argon time projection chamber (LArTPC) detectors through the Short Baseline Neutrino (SBN)~\cite{bib:SBN} program and Deep Underground Neutrino Experiment (DUNE)~\cite{bib:DUNE}. 
%These experimental setups are also ideal for searches for new physics beyond the Standard Model (SM) performed leveraging Fermilab's intense neutrino beamlines. 
%These precise measurements of neutrino oscillation parameters will be significantly impacted by uncertainties in modeling the neutrino interaction rate on argon. Similarly, neutrino interactions constitute a background for beyond the standard model (BSM) processes.  
Uncertainties in modeling the neutrino interaction rate on argon impact the precision to which neutrino oscillation parameter measurements can be performed. Similarly, neutrino interactions constitute a background for beyond the standard model (BSM) processes~\cite{bib:bsmxsec}.  
In both cases, accurate modeling of the interaction rate and final-state particles produced in neutrino interactions is a crucial part of this experimental program. This has led to a broad program focused on studying neutrino interactions to support and enhance the upcoming neutrino oscillation and BSM physics programs~\cite{bib:SNOWMASS_NF6_REPORT}. %Measurements of neutrino interactions are a key component of this experimental effort. 
%the experimental program that aims to perform precision neutrino oscillation measurements and searches for BSM physics with neutrino beams in the sub-GeV to few-GeV energy range.

Neutrinos interact with atomic nuclei with a broad range of interaction modes. An important process in the $\mathcal{O}$(GeV) energy range is resonant interactions (RES) where a neutrino strikes a single nucleon (neutron or proton) exciting a baryon resonance. Uncertainties on the modeling of these processes contribute to the overall systematics on neutrino event rates. RES interactions and their modeling uncertainty play a particularly important role in both short- and long-baseline experiments due to the production of final-states which mimic signatures of $\nu_{\mu}\rightarrow\nu_e$ oscillation and BSM observables. Constraints on resonant interactions, particularly on argon, can contribute to validations and improvements of such interaction models. Moreover, resonant interactions are one of the dominant interaction modes for the long-baseline DUNE neutrino experiment. 

A broad category of baryon resonances can be excited when neutrinos strike a nucleon~\cite{bib:ReinSehgal,bib:REStheory}.
Most resonances decay to a nucleon and a charged or neutral pion, and this final state has been the most frequently studied to-date in RES neutrino-nucleus interactions. These interactions are dominated by the excitation and decay of the $\Delta(1232)$ resonant state. However, higher order resonances, while subdominant, contribute at the $\sim$10\% level to the total event rate. If not properly accounted for, these resonances can lead to mis-modeled backgrounds in precision oscillation measurements and BSM searches. Yet, testing their modeling in neutrino interactions is made difficult by the lack of experimental measurements.

Resonances such as the $N(1535)$, $N(1650)$, and $N(1710)$ states have sizeable (though with large uncertainties) branching fractions to $\eta$ production of 30-55\%, 15-35\%, and 10-50\%, respectively~\cite{bib:PDG}. \textcolor{black}{For context, roughly 1-2\% of all neutrino interactions in DUNE will lead to $\eta$ mesons in the final state.} Measuring $\eta$ production in neutrino interactions is a promising way to study RES interactions targeting resonant states that cannot be easily probed through measurements of pion production.
%Such baryon resonances can also decay producing an $\eta$, though often do so with negligible branching ratios. However, resonances such as the $N(1535)$, $N(1650)$ and $N(1710)$ have sizeable (though with large uncertainties) branching ratios for $\eta$ production of 30-55\%, 15-35\%, and 10-50\%, respectively~\cite{bib:PDG}. 
%This makes the measurement of $\eta$ production in neutrino interactions a complementary way of studying RES interactions, targeting resonant states that cannot be easily probed through measurements of pion production. 
The BEBC WA59 collaboration reported a measurement of $\eta$ production on a Ne-H${}_{2}$ target~\cite{bib:BEBCWA59}, and 13 candidate $\eta$ events were seen by the ICARUS experiment operating at LNGS in an unpublished study~\cite{bib:ICARUS_eta}. Both measurements were performed in the multi-GeV neutrino beams of the SPS at CERN. %The measurement reported here represents the first extraction of the cross-section for $\eta$ production on argon, and the first measurement in a $\sim$GeV energy beam, with mean energy of $<E_{\nu}> \sim 0.8$ GeV. 
Theoretical calculations for the cross section for $\eta$ production in neutrino interactions are reported in Refs.~\cite{bib:etatheory1,bib:etatheory2,bib:etatheory3}.
%provides a valuable new benchmark for neutrino-argon interaction modeling.

We present the first measurement of the cross section for $\eta$ production in neutrino interactions on argon. The measurement uses 6.79 $\times 10^{20}$ protons on target (POT) of neutrino data collected on-axis on the Booster Neutrino Beamline (BNB)~\cite{bib:bnb} by MicroBooNE during the first three years of operation, 2016 to 2018. The analysis leads to  the largest sample of $\eta$ meson candidates observed in neutrino-argon interactions and is the first measurement of their production on any target in a beam of sub-GeV mean energy. Being the first quantitative measurement of $\eta$ production on argon, this measurement opens a completely new area for probing neutrino interactions.

In addition to the important impact on cross section modeling, the ability to observe $\eta$ decays in a LArTPC can find broader application. We identify three additional ways in which $\eta$ particle measurements in LArTPCs can have a significant impact on neutrino, nuclear, and BSM physics searches:
\begin{enumerate}
    \item The ability to observe $\eta$ decays in a LArTPC opens the door for searches of proton-decay in the $p\rightarrow e^{+} + \eta$ and $p\rightarrow \mu^{+} + \eta$ channels with the DUNE experiment. This is a channel that has already been used for proton-decay searches by Super-K~\cite{bib:superk_pdecay} with competitive limits of $\sim10^{34}$ years. This decay channel complements the primary focus of DUNE on the $K^{+} + \bar{\nu}$ decay mode.
    \item Measurements of $\eta$ particles through their decay to photon pairs provide a novel tool for the calibration of the electromagnetic (EM) energy scale, a critical component of the $\nu_e$ lepton energy determination for the extraction of $\delta_{\textrm {CP}}$ and other oscillation parameters. Decays to photon pairs from $\eta$ particles provide a sample of higher energy showers which complement the $\mathcal{O}$(50-200 MeV) photons from $\pi^0$ decay~\cite{bib:uB_pi0}. Photons from $\eta$ decay, in particular, have greater overlap with the energy of electrons expected from the $\nu_e$ flux component of  SBN \textcolor{black}{and will allow for a data-driven validation of shower energy-scale reconstruction linearity up to GeV energies.}
    \item Finally, the large uncertainty in current experimental measurements of baryon resonance decays to the $\eta$~\cite{bib:PDG} can be constrained through precise measurements of $\eta$ production in neutrino interactions.
\end{enumerate}
These items indicate the large impact this and future measurements of $\eta$ production in a LArTPC can have across different areas of particle physics.

%\textbf{MicroBooNE} 
The MicroBooNE detector~\cite{bib:uB_detector} comprises a TPC with 85 tonnes of liquid argon active mass accompanied by a photon detection system made up of 32 photomultiplier tubes (PMTs). Neutrino interactions on the argon target are recorded through the ionization and scintillation light signatures produced by final-state charged particles traversing the detector volume. Ionization charge is recorded on three wire planes allowing the experiment to obtain millimeter-resolution three-dimensional images of neutrino interactions. Scintillation light collected on the PMT array provides the timing resolution necessary to identify neutrino interactions in-time with the BNB and to reject cosmic-ray backgrounds.

%\textbf{Simulation and Reconstruction} 
The simulation of neutrino interactions and particle propagation through the MicroBooNE detector is carried out within the LArSoft framework~\cite{bib:larsoft}. The BNB neutrino flux at the MicroBooNE detector is simulated leveraging the flux simulation developed by the MiniBooNE collaboration~\cite{bib:mbflux} accounting for MicroBooNE's position along the beamline. Neutrino interactions in the detector are simulated with the \texttt{GENIE v3.0.6 (G18\_10a\_02\_11a)} event generator~\cite{bib:GENIEv3} that was tuned to CC0$\pi$ data from the T2K collaboration~\cite{bib:t2ktunedata} as described in Ref.~\cite{bib:uB_genietune}.  Resonances are modeled according to the description of Rein and Sehgal~\cite{bib:ReinSehgal} and are allowed to decay based on tabulated branching ratios from the Particle Data Group~\cite{bib:PDG}. Decays of resonances above the $\Delta(1232)$ are treated as isotropic. While multiple resonances can contribute to $\eta$ production, only a few do so at a meaningful rate.  \textcolor{black}{In particular, the $N(1535)$ is predicted to contribute the dominant rate of $\eta$ production (87\%) according to the GENIE generator simulation used in this analysis. It is important to note however that the GENIE simulation does not account for interference between the different resonances, and is further subject to the large uncertainty in the branching fractions of these resonant states}. While based on simulation, this observation suggests that studies of $\eta$ production in the BNB can serve as a unique selector of a pure sample of events from a single non-$\Delta$ resonant state. This provides new handles for detailed studies and model constraints for RES interactions.

%Within this generator, $\eta$ production is modeled using the \textcolor{red}{XXX} model.
Particle propagation through the detector is carried out via the \texttt{GEANT4} simulation~\cite{bib:geant4}, and propagation of ionization and scintillation signals is carried out through dedicated algorithms that model the detector's response. Simulated neutrino interactions are overlayed with data events collected with an unbiased trigger in anti-coincidence with the beam which allows for data-driven cosmic-ray and detector noise modeling.
PMT signals from MicroBooNE's data are used to apply an online trigger that rejects events with little visible light collected in coincidence with the 1.6 $\mu$s BNB neutrino spill. Offline, PMT signals from both data and simulation are processed through reconstruction algorithms that measure the photo-electrons (PE) on each PMT associated to the interaction in-time with the BNB spill.
Both data and simulated events undergo the same reconstruction workflow. Noise filtering~\cite{bib:uB_noise} and signal-processing~\cite{bib:uB_signal1,bib:uB_signal2} algorithms are applied to TPC wire signals to measure energy deposits on each wire-plane. The \texttt{Pandora} multi-algorithm pattern recognition framework~\cite{bib:pandora} is used to reconstruct three-dimensional particle trajectories and a particle flow hierarchy and to identify the $\mathcal{O}$(10) interactions (mostly cosmic-rays) occurring in each recorded event. 
MicroBooNE's TPC and PMT signals are calibrated to account for position and time-dependent variations in detector response. PMT gains are calibrated for each PMT independently on a weekly basis, and the overall light yield in the detector is calibrated through a single time-dependent correction factor. MicroBooNE's TPC signal calibration accounts for position and time-dependent variations in the detector's ionization production, transport, and signal formation. These calibrations account for the variation in the detector's position-dependent electric field~\cite{bib:uB_SCE1,bib:uB_SCE2} and for the relative and absolute charge-scale calibration~\cite{bib:uB_TPCcalib}. 
Electromagnetic (EM) shower energy calibration is performed through the methods described in Ref.~\cite{bib:uB_pi0} leading to a shower energy correction of $\times 1.20$ to account for energy deposited by the shower not collected by the reconstruction. % a definition of shower energy of $E_{\textrm{reco}} = E_{\textrm{calorimetry}}/0.$.
The calibration of the detector's calorimetric response is particularly relevant to this analysis which relies on calorimetry to measure the energy of EM showers.

%Neutrinos can interact with atomic nuclei in a broad range of interaction modes, from deep inelastic scattering (DIS) off of individual quarks in a nucleon, to coherent scattering off the entire nucleus. A dominant interaction mode in the $\mathcal{O}$(1) GeV energy range involves interactions with a single nucleon which excites a resonance (RES). A broad category of resonances can be excited, with the . Most resonances decay to a nucleon and a charged or neutral pion, and the pion final-state is the channel that is most studied in neutrino-nucleus interactions [cite]. Excited resonances can decay to $\eta$ particles, though often do so with negligible branching ratios. The N(1535), N(1650) and N(1710) resonances have the largest expected branching ratios for $\eta$ production: 0.40-0.55, 0.15-0.35, and 0.10-0.50 respectively~\cite{PDG}. These branching ratios, measured in XXX experiments, have a significant degree of uncertainty. The dominant channels for $\eta$ particle decay are $\eta \rightarrow 2\gamma$ (40\%), $\eta \rightarrow 3\pi^0$ (33\%), and $\eta \rightarrow \pi^0 + \pi^+ + \pi^-$ (23\%). This analysis targets the decay to two photons due to the clean final-state signature and large branching ratio.

%\textbf{Simulation of $\eta$ Mesons} 
The $\eta$ meson has multiple decay modes with comparable branching fractions. 
%This requires a careful signal definition to account for the impact of different decay modes and final-state interactions (FSI) that may alter the visible particles in the detector.
The dominant channels are $\eta \rightarrow 2\gamma$, $\eta \rightarrow 3\pi^0$, and $\eta \rightarrow \pi^0 + \pi^+ + \pi^-$, with branching ratios of 40\%, 33\%, and 23\%, respectively~\cite{bib:PDG}. This analysis targets the decay to two photons given that it is the dominant decay mode, and it leads to the cleanest final-state signature. The very low rate expected for $\eta$ production in MicroBooNE ($<$ 1\% of all $\nu$ interactions) makes the $2\gamma$ signature particularly attractive due to the powerful background rejection that can be achieved by selecting for a $2\gamma$ invariant mass consistent with 548 MeV/$c^2$, the mass of the $\eta$ meson.
The signal for this analysis is defined as events in which an $\eta$ particle is produced as a result of the neutrino-argon interaction and where there are two photons and no $\pi^0$ present in the final-state. \textcolor{black}{No other activity from charged particles at the vertex is required to identify the candidate event.} While muon neutrinos make up $\sim$95\% of the BNB flux, neutrinos and anti-neutrinos of all flavors are included in the signal definition. Finally, this analysis does not apply selection cuts on the presence of an outgoing lepton in the interaction and, therefore, targets $\eta$ production from both charged current (CC) and neutral current (NC) processes. The interaction process being sought can therefore be described as $\nu_{\textrm{CC} + \textrm{NC}} \rightarrow \eta + 0\pi^0 + X \rightarrow 2\gamma + 0\pi^0 + X$ with $X$ denoting any additional particles of any multiplicity.

%\textbf{Event Selection} 
Neutrino interactions are identified using both scintillation light and TPC signals. Interactions which are out-of-time with respect to the in-time TPC drift window are rejected. Remaining TPC interactions which are inconsistent with the in-time scintillation light signal collected by the PMTs are discarded. At this stage, a comparable rate of selected neutrino to cosmic-ray interactions is achieved with partially-contained in-time cosmic-ray interactions comprising the bulk of selected backgrounds. This yields an 83\% efficiency for identifying neutrino interactions.

After isolating neutrino interactions, cuts are applied to isolate the $2\gamma$ topology being sought. The selection is implemented leveraging the tools developed in Ref.~\cite{bib:uB_PeLEE}. Neutrino candidates are required to have an interaction vertex in the TPC fiducial volume and a \texttt{Pandora} topological neutrino score greater than 0.1~\cite{bib:Wouter}. Diphoton events are selected by requiring exactly two reconstructed showers with greater than 50 MeV of reconstructed energy in each shower. The requirement that exactly two showers are reconstructed serves to reject events with an $\eta$ and additional $\pi^0$ as well as events where the $\eta$ decays via the three $\pi^0$ mode. Two quality cuts are further applied to reconstructed showers: a minimum distance from the reconstructed neutrino interaction vertex of 2~$\mbox{cm}$ is required and showers must have a reconstructed direction that is aligned with the direction connecting the shower to the interaction vertex ($\cos\theta_{\textrm{shower}} > 0.9$). At this stage the selection efficiency is 19.5\% and the purity 3.5\% with backgrounds dominated by $\pi^0$ events.

To reject $\pi^0$ events and select $\eta$ candidates, events with a diphoton mass smaller than 250 MeV/$c^2$ and larger than 750 MeV/$c^2$ are rejected. This requirement brings the efficiency to 18.2\% with a one order of magnitude increase in purity (30.2\%). Diphoton pairs from $\pi^0$ candidates are used to validate and refine the energy scale calibration for EM showers leading to an additional energy scale correction of 5.2\%~\cite{bib:supplementary}. %The study justifying this correction is presented in the supplementary material~\cite{bib:supplementary}.

\begin{figure}[ht]
\begin{center}
\includegraphics[width=0.48\textwidth]{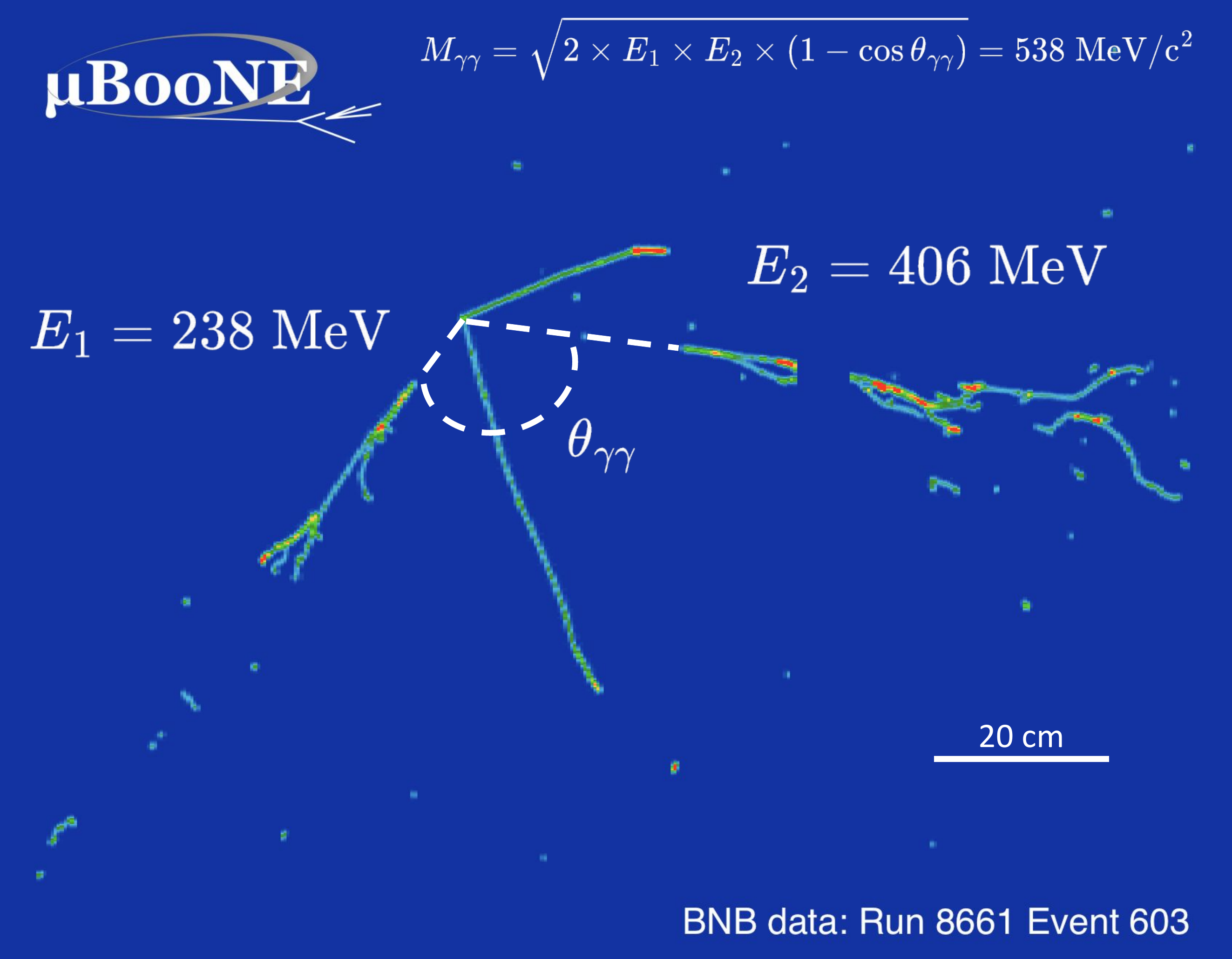}%{images/eta_EVD.pdf}
\caption{\label{fig:evd} Event display of candidate $\eta$ event.}
\end{center}
\end{figure}

Residual backgrounds consist of mis-reconstructed $\pi^0$ events and interactions with two or more $\pi^0$s in the final state. These residual backgrounds are rejected by relying on the kinematics of the $\eta \rightarrow 2\gamma$ decay. 
Given two neutral particles of different mass but equivalent total energy decaying to two photons, the lighter particle will produce a more highly boosted diphoton pair. 
To leverage this kinematic constraint, we require that selected diphoton pairs have an opening angle such that $\cos\theta_{\gamma\gamma} < 0.5$.
%For a particle of mass $M$ and total energy $E$ decaying to two photons, the opening angle is kinematically constrained to be larger than a certain value $\theta_{\textrm{min}}$, given by
%\begin{equation}
%    \label{eq:minangle}
%    \sin\left(\frac{\theta_{\textrm{min}}}{2}\right) = \frac{M}{E}.
%\end{equation}
The $2\gamma$ decay allows us to define a kinematically minimal mass %for the diphoton pair,
for a diphoton pair with minimum opening angle $\theta_{\gamma\gamma}$,
%Equation~\ref{eq:minangle} can further be inverted to obtain the reconstructed mass threshold
\begin{equation}
    \label{eq:minmax}
    M_{\textrm{max}} = E_{\gamma\gamma} \sqrt{\frac{1}{2}\left(1-\cos\theta_{\gamma\gamma}\right)},
\end{equation}
where $E_{\gamma\gamma}$ is given by the sum of the energy of the two photons. This quantity provides a powerful discriminant for particles of different mass and relies only on the opening angle between the two photons and the sum of the shower energies. Therefore, the dependence on the accuracy of the reconstructed energy for each individual shower is reduced. A cut requiring that events have a value of $M_{\textrm{max}} > 400$ MeV/$c^2$ is applied bringing the final selection purity and efficiency to 49.9\% and 13.6\%, respectively. Importantly, while relying on event kinematics, this cut is tailored to cause minimal bias in selecting signal events leading to a flat efficiency for $\eta$ particles with energies in the 0.5-1.0 GeV range. %More details on the motivation for and impact on the selection of the final two kinematics-based cuts
Distributions for $\cos\theta_{\gamma\gamma}$ and $M_{\textrm{max}}$ which show the separation between signal and background achieved through the use of these variables are provided in the supplementary material~\cite{bib:supplementary}.
%This concept is conveyed by the expression for the  By rejecting events with $\cos\theta_{\gamma\gamma} > 0.5$ and XXX a selection purity of 49.9\% is achieved. The selection efficiency is found to be $13.6 \pm 0.3$\%, and is flat in the $\eta$ energy range of 0.5-1.0 GeV. 
A total of 93 events are selected in the dataset used in this analysis. A candidate $\eta$ event from this dataset is shown in Fig.~\ref{fig:evd}. While the analysis is inclusive of CC and NC processes, the selection is dominated by CC interactions according to the simulated prediction. This is a consequence of the larger relative content of 3:1 for CC:NC events in the simulation as well as a larger selection efficiency for CC $\eta$ production \textcolor{black}{(15.4\%, compared to 8.9\% for NC)}. Dedicated measurements of NC and CC $\eta$ production will be pursued in future work.

%\textbf{Cross Section Extraction} 
This analysis measures a single-bin, flux-integrated cross section for $\eta$ production. The measurement is carried out by calculating the expression
\begin{equation}
    \label{eq:xsec}
    \sigma = \frac{N - B}{\epsilon \times N_{\rm target} \times \Phi_{\nu}},
\end{equation}
with $N$ and $B$ the selected number of data events and expected number of background events, respectively, $\epsilon$ the efficiency for signal events (13.6\%), $N_{\rm target}$ the number of target nucleons ($4.057\times 10^{31}$), and $\Phi_{\nu}$ the integrated neutrino flux ($5.01 \times 10^{11} \; \nu/\textrm{cm}^2$). Backgrounds from $1\pi^0$ and multi-$\pi^0$ events are constrained in a data-driven way to improve the accuracy and to reduce the overall uncertainty on the extracted $\eta$ production cross section. The supplementary material describes how this constraint is carried out~\cite{bib:supplementary}.
%\begin{equation}
%\label{eq:systematics00}
%sigma^2_{\rm multi-sim} &=& \frac{1}{N} \sum^{N}_{k=1} %\left(n^{\rm CV} - n^{k} \right)^2 
%\end{equation}
\textcolor{black}{A fake-data study is performed using events generated via the \texttt{NuWro} event generator treated as data. The fake-data study included the full sideband constraint procedure and led to an extracted cross section within 1$\sigma$ of the NuWro truth value.}

\begin{figure}[ht]
\begin{center}
\includegraphics[width=0.48\textwidth]{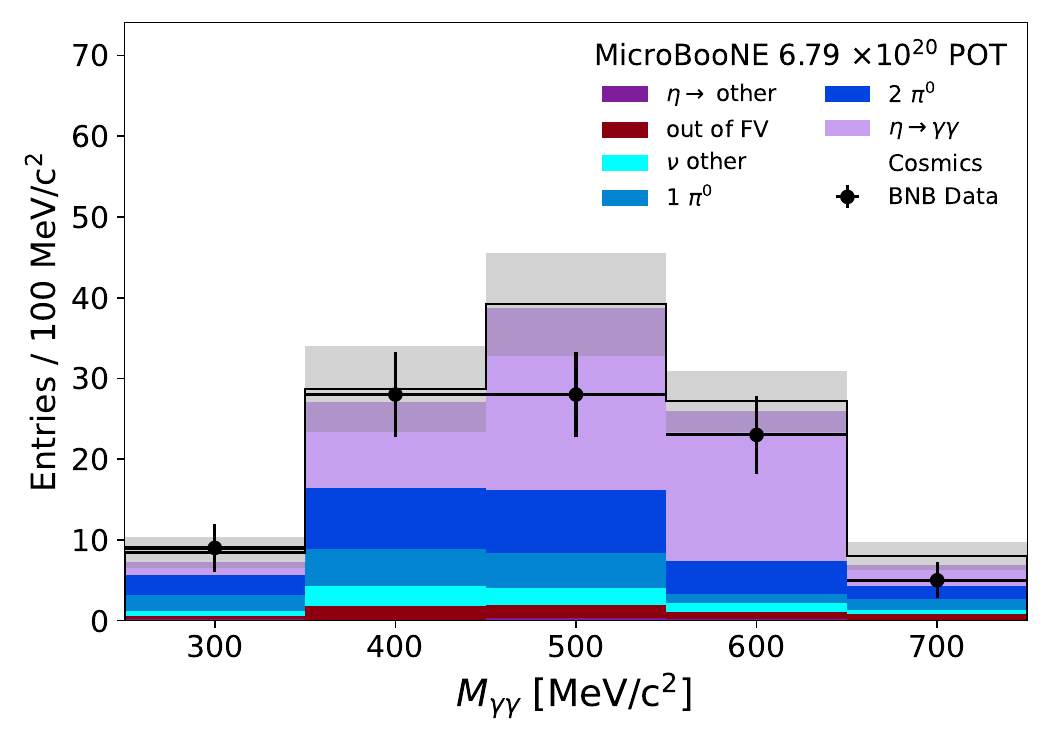}
\caption{\label{fig:etamass} Distribution of $M_{\gamma\gamma}$ for selected $\eta$ candidates showing data (data points with statistical uncertainties denoted by the error bar) and the predicted event rate (stacked histogram). Different colors denote different topologies, as described in the legend. The gray error band denotes the systematic uncertainty on the predicted event rate.}
\end{center}
\end{figure}

Figure~\ref{fig:etamass} shows the distribution of $M_{\gamma\gamma}$ for $\eta$ candidates after applying the full event selection. 
%Solid and hashed stacked histograms denote predicted events, including constrained $1\pi^0$ and multi-$\pi^0$ events (light and dark blue, respectively). Purple denotes the predicted signal rate for $\eta$ particles decaying to two photons and 0$\pi^0$. The small darker purple band denotes true $\eta$ events which decay via other modes. The distributions in cyan, brown, and hashed white represent background contributions from other neutrino interactions in the TPC, neutrino interactions occurring outside of the fiducial volume, and cosmic-ray induced events, respectively. 
The simulated prediction (stacked histogram in Fig.~\ref{fig:etamass}) shows a peak for the signal sample in the 450-550 MeV/$c^2$ bin consistent with the $\eta$ mass of 548 MeV/$c^2$. 

Systematic uncertainties for the measurement are assessed by studying the impact of model variations on the extracted cross section. The constrained uncertainty due to modeling of the neutrino flux, cross section model, and particle re-interactions in the detector leads to an uncertainty of 14.2\% for $1\pi^0$ and multi-$\pi^0$ events. % obtained through the constraint procedure described in the supplementary material~\cite{bib:supplementary}. 
This uncertainty to the cross section contributed by non-$\pi^0$ backgrounds is found to be 10.4\% and is left unconstrained. As detailed in Ref.~\cite{bib:uB_wiremod}, detector systematic uncertainties account for discrepancies between data and simulation in charge and light response. Detector modeling leads to a 17.7\% systematic uncertainty on the extracted cross section. Additional uncertainties on the extracted cross section are due to simulation sample statistics (7.6\%), uncertainty on the number of argon targets (1.0\%), POT exposure (2.0\%), and \textcolor{black}{the impact of sample statistics on the} selection efficiency (2.0\%). The total systematic uncertainty is calculated to be 26.3\%. The data statistical uncertainty is 25.6\%.
\textcolor{black}{While this analysis reports a cross section inclusive of CC and NC interactions, we highlight the differences in efficiency for these two channels and the magnitude of systematic uncertainties on their modeled ratio and efficiency. The efficiencies for CC and NC interactions are $14.3 \pm 2.8$\% and $8.9 \pm 0.4$\%, respectively, where uncertainties denote the uncertainty due to cross section model variations. The selection efficiency, including all systematic uncertainties, is $13.6 \pm 2.4$\%. Finally, the cross section modeling uncertainty on the predicted CC to NC ratio is 20\%. The impact of these uncertainties will be meaningful in future high statistics measurements.}

\begin{comment}
\renewcommand{\arraystretch}{1.5}
\begin{table}[ht]
\begin{center}
\begin{tabular}{{|p{0.23\textwidth}|p{0.23\textwidth}|}}
\hline
\multicolumn{2}{|c|}{\textbf{Cross Section Uncertainties}} \\
\hline
source of uncertainty & uncertainty [\%] \\
\hline
non-$\pi^0$ backgrounds & 10.4 \\
$\pi^0$ backgrounds & 15.0 \\
detector & 18.3 \\
efficiency & 2.0 \\
MC statistics & 7.6 \\
\hline
total systematic & 27.9 \\
data statistics & 26.3 \\
\hline
total uncertainty & 37.7 \\
\hline
\end{tabular}
\end{center}
\caption{\label{tab:xsecerr}Cross Section Uncertainty,}
\end{table}
\renewcommand{\arraystretch}{1.0}
\end{comment}

The measured cross section per nucleon for a final-state with two photons and no $\pi^0$ in the final state tagged by the selection is found to be $\sigma_{\nu\rightarrow 1\eta + X \rightarrow 2\gamma + 0\pi^0 + X} = 1.27 \pm 0.33 \; \textrm{(stat.)} \pm 0.34 \; \textrm{(syst.)} \; 10^{-41} {\rm cm}^{2} / \textrm{nucleon}$.
%\begin{widetext}
%\begin{equation}
%\label{eq:xsec}
%\sigma_{\nu\rightarrow 1\eta + x \rightarrow \gamma\gamma + 0\pi^0 + x} = 1.27 \pm 0.33 \; \textrm{(stat.)} \pm 0.34 \; \textrm{(syst.)} \; 10^{-41} {\rm cm}^{2} / \textrm{nucleon}
%\end{equation}
%\end{widetext}
Due to its $>10^{-19}$ second lifetime, the $\eta$ decays almost always outside of the struck nucleus, and \textcolor{black}{while} final-state interactions \textcolor{black}{can affect the propagation of the $\eta$ particle as it exits the nucleus, they do not} impact the particular decay mode chosen. The measured cross section can then be corrected for the well measured $\eta$ branching ratio to two photons of $39.41\% \pm 0.20\%$~\cite{bib:PDG}. This leads to a total cross section for $\eta$ production ($\sigma_{\nu\rightarrow 1\eta + X}$) of $3.22 \pm 0.84 \; \textrm{(stat.)} \pm 0.86 \; \textrm{(syst.)} \; 10^{-41} {\rm cm}^{2} / \textrm{nucleon}$.
The reported cross section is integrated over all contributions to the MicroBooNE flux from $\nu_{\mu}$ (93.7\%), $\bar{\nu}_{\mu}$ (5.8\%), $\nu_e$ (0.5\%), and $\bar{\nu}_e$ (0.05\%). In simulation, 98.6\% of selected signal events originate from $\nu_{\mu}$ interactions, 0.9\% from $\bar{\nu}_{\mu}$, and 0.5\% from $\nu_e$.

The extracted cross section ($\sigma_{\nu\rightarrow 1\eta + X}$) can be compared to that for different neutrino interaction generators. For the \texttt{GENIE} generator, a cross section of $4.63$ and $4.61\times10^{-41} {\rm cm}^{2} / \textrm{nucleon}$ is calculated for this signal definition for the \texttt{GENIE v2\_12\_10} and \texttt{GENIE v3\_00\_06 G18\_10a\_02\_11a} models respectively. The \texttt{NuWro 19.02.1}~\cite{bib:nuwro} generator gives a cross section of $5.45\times10^{-41} {\rm cm}^{2} / \textrm{nucleon}$, and \texttt{NEUT v5.4.0}~\cite{bib:NEUT} gives a cross section of $11.9\times10^{-41} {\rm cm}^{2} / \textrm{nucleon}$. Both versions of \texttt{GENIE}, as well as \texttt{NuWro}, give a cross section which is larger than observed but still within $1-2\sigma$ of the measured value accounting for uncertainties. The \texttt{NEUT} cross section is found to be significantly larger than what is observed in data. The supplementary material shows a figure comparing the data result to various generator predictions~\cite{bib:supplementary}.

\begin{figure}[h]
\begin{center}
\includegraphics[width=0.48\textwidth]{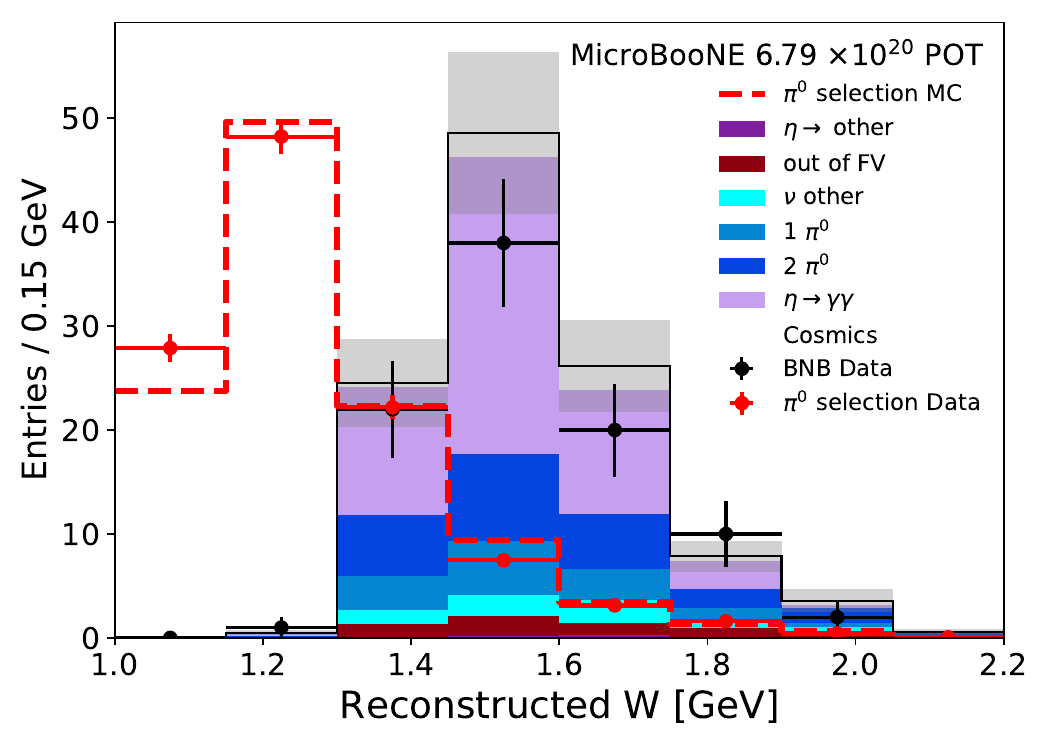}
\caption{\label{fig:recow} Reconstructed invariant mass of the hadronic system utilizing the four-momenta of the reconstructed $\eta$ and leading proton (if identified) in the event. The black solid line and data points show the distribution for $\eta$ candidate events predicted and observed, respectively. The distributions in red show the same reconstructed quantity for events from the MicroBooNE data compared to prediction from the $\pi^0$ sideband, normalized to the same number of events from the prediction for the $\eta$ selection.}
\end{center}
\end{figure}

The sample of $\eta$ candidate events is additionally employed to reconstruct the invariant mass of the hadronic system to probe the excited resonance. This is calculated using additional information from the hadronic system produced in the interaction.
If protons are identified as exiting the neutrino vertex, then the leading proton is combined with the 4-vector of the $\eta$ to calculate the mass $W$ of the hadronic system. 
%In addition to the 4-vector of the $\eta$ candidate, that of the leading reconstructed proton exiting the neutrino vertex is included in the calculation of $W$, if one is identified. 
Protons are identified through the particle identification methods presented in Ref.~\cite{bib:uB_PID}. The reconstructed $W$ is shown in Fig.~\ref{fig:recow} for the events selected by the analysis. 

The data and simulation show good agreement, and the distribution peaks at $\sim$1.5 GeV in agreement with the expectation that most $\eta$ particles are produced though an excitation of the $N(1535)$ resonance. %The red dashed histogram and red data points show the same distribution reconstructed with events from the $1\pi^0$ selection but area-normalized to the simulation prediction for the $\eta$ selection. 
In absolute terms, there are over one order of magnitude more $\pi^0$ candidates than selected $\eta$ candidate events. The $\pi^0$ dominated distribution shows a clear separation from that for $\eta$ candidates, peaking at $\sim$1.2 GeV as expected for events produced through an excitation of the $\Delta(1232)$ resonance. \textcolor{black}{Isolating $\eta$ production events allows to suppress the large rate of $\Delta(1232)$ events which would otherwise swamp higher resonances making their study challenging.} This represents the first demonstration of the ability to identify higher-order resonances other than the $\Delta(1232)$ in neutrino-nucleus interactions and provides a new powerful tool for the study of RES interactions.

%\textbf{Conclusions} 
In summary, this letter presents the first cross section measurement of $\nu$-Ar $\eta$ production. Future measurements of $\eta$ production in MicroBooNE will benefit from additional data for higher statistics measurements. The measurement of $\eta$ production in LArTPCs launched through this work will further flourish with the SBND~\cite{bib:SBN_REVIEW} and DUNE-ND~\cite{bib:DUNE_ND} detectors which will leverage significantly larger neutrino flux in order to report results with $\gtrsim 10^3$ candidate events. These will have a significant impact on measurements of resonant interaction processes and, in particular, a unique ability to constrain higher-order resonances above the $\Delta(1232)$ \textcolor{black}{up to uncertainties in their branching ratios}. \textcolor{black}{Future high statistics cross section measurements of $\eta$ production will nonetheless have to confront challenges in constraining the sizeable single- and multi-$\pi^0$ background processes which are subject to large modeling uncertainties, with particular attention needed in how sideband constraints are used to extrapolate background predictions into the signal region.} In addition, these samples will provide a new tool for the calibration of EM showers that are of particular importance to the oscillation and BSM physics programs that are being carried out with these detectors.

This document was prepared by the MicroBooNE collaboration using the resources of the Fermi National Accelerator Laboratory (Fermilab), a U.S. Department of Energy, Office of Science, HEP User Facility. Fermilab is managed by Fermi Research Alliance, LLC (FRA), acting under Contract No. DE-AC02-07CH11359. MicroBooNE is supported by the following: the U.S. Department of Energy, Office of Science, Offices of High Energy Physics and Nuclear Physics; the U.S. National Science Foundation; the Swiss National Science Foundation; the Science and Technology Facilities Council (STFC), part of the United Kingdom Research and Innovation; the Royal Society (United Kingdom); and the UK Research and Innovation (UKRI) Future Leaders Fellowship. Additional support for the laser calibration system and cosmic ray tagger was provided by the Albert Einstein Center for Fundamental Physics, Bern, Switzerland. We also acknowledge the contributions of technical and scientific staff to the design, construction, and operation of the MicroBooNE detector as well as the contributions of past collaborators to the development of MicroBooNE analyses, without whom this work would not have been possible. For the purpose of open access, the authors have applied a Creative Commons Attribution (CC BY) license to any Author Accepted Manuscript version arising from this submission.

\bibliography{zapssamp}% Produces the bibliography via BibTeX.

%merlin.mbs apsrev4-1.bst 2010-07-25 4.21a (PWD, AO, DPC) hacked
%Control: key (0)
%Control: author (8) initials jnrlst
%Control: editor formatted (1) identically to author
%Control: production of article title (-1) disabled
%Control: page (0) single
%Control: year (1) truncated
%Control: production of eprint (0) enabled
\providecommand{\noopsort}[1]{}\providecommand{\singleletter}[1]{#1}%
\begin{thebibliography}{38}%
\makeatletter
\providecommand \@ifxundefined [1]{%
 \@ifx{#1\undefined}
}%
\providecommand \@ifnum [1]{%
 \ifnum #1\expandafter \@firstoftwo
 \else \expandafter \@secondoftwo
 \fi
}%
\providecommand \@ifx [1]{%
 \ifx #1\expandafter \@firstoftwo
 \else \expandafter \@secondoftwo
 \fi
}%
\providecommand \natexlab [1]{#1}%
\providecommand \enquote  [1]{``#1''}%
\providecommand \bibnamefont  [1]{#1}%
\providecommand \bibfnamefont [1]{#1}%
\providecommand \citenamefont [1]{#1}%
\providecommand \href@noop [0]{\@secondoftwo}%
\providecommand \href [0]{\begingroup \@sanitize@url \@href}%
\providecommand \@href[1]{\@@startlink{#1}\@@href}%
\providecommand \@@href[1]{\endgroup#1\@@endlink}%
\providecommand \@sanitize@url [0]{\catcode `\\12\catcode `\$12\catcode
  `\&12\catcode `\#12\catcode `\^12\catcode `\_12\catcode `\%12\relax}%
\providecommand \@@startlink[1]{}%
\providecommand \@@endlink[0]{}%
\providecommand \url  [0]{\begingroup\@sanitize@url \@url }%
\providecommand \@url [1]{\endgroup\@href {#1}{\urlprefix }}%
\providecommand \urlprefix  [0]{URL }%
\providecommand \Eprint [0]{\href }%
\providecommand \doibase [0]{http://dx.doi.org/}%
\providecommand \selectlanguage [0]{\@gobble}%
\providecommand \bibinfo  [0]{\@secondoftwo}%
\providecommand \bibfield  [0]{\@secondoftwo}%
\providecommand \translation [1]{[#1]}%
\providecommand \BibitemOpen [0]{}%
\providecommand \bibitemStop [0]{}%
\providecommand \bibitemNoStop [0]{.\EOS\space}%
\providecommand \EOS [0]{\spacefactor3000\relax}%
\providecommand \BibitemShut  [1]{\csname bibitem#1\endcsname}%
\let\auto@bib@innerbib\@empty
%</preamble>
\bibitem [{\citenamefont {Antonello}\ \emph {et~al.}(2015)\citenamefont
  {Antonello} \emph {et~al.}}]{bib:SBN}%
  \BibitemOpen
  \bibfield  {author} {\bibinfo {author} {\bibfnamefont {M.}~\bibnamefont
  {Antonello}} \emph {et~al.} (\bibinfo {collaboration} {MicroBooNE, LAr1-ND,
  ICARUS-WA104}),\ }\href@noop {} {\  (\bibinfo {year} {2015})},\ \Eprint
  {http://arxiv.org/abs/1503.01520} {arXiv:1503.01520 [physics.ins-det]}
  \BibitemShut {NoStop}%
\bibitem [{\citenamefont {Abi}\ \emph {et~al.}(2020)\citenamefont {Abi} \emph
  {et~al.}}]{bib:DUNE}%
  \BibitemOpen
  \bibfield  {author} {\bibinfo {author} {\bibfnamefont {B.}~\bibnamefont
  {Abi}} \emph {et~al.} (\bibinfo {collaboration} {DUNE}),\ }\href@noop {} {\
  (\bibinfo {year} {2020})},\ \Eprint {http://arxiv.org/abs/2002.03005}
  {arXiv:2002.03005 [hep-ex]} \BibitemShut {NoStop}%
\bibitem [{\citenamefont {Coyle}\ \emph {et~al.}(2022)\citenamefont {Coyle},
  \citenamefont {Li},\ and\ \citenamefont {Machado}}]{bib:bsmxsec}%
  \BibitemOpen
  \bibfield  {author} {\bibinfo {author} {\bibfnamefont {N.~M.}\ \bibnamefont
  {Coyle}}, \bibinfo {author} {\bibfnamefont {S.~W.}\ \bibnamefont {Li}}, \
  and\ \bibinfo {author} {\bibfnamefont {P.~A.~N.}\ \bibnamefont {Machado}},\
  }\href {\doibase 10.1007/JHEP12(2022)166} {\bibfield  {journal} {\bibinfo
  {journal} {JHEP}\ }\textbf {\bibinfo {volume} {12}},\ \bibinfo {pages} {166}
  (\bibinfo {year} {2022})},\ \Eprint {http://arxiv.org/abs/2210.03753}
  {arXiv:2210.03753 [hep-ph]} \BibitemShut {NoStop}%
\bibitem [{\citenamefont {Balantekin}\ \emph {et~al.}(2022)\citenamefont
  {Balantekin} \emph {et~al.}}]{bib:SNOWMASS_NF6_REPORT}%
  \BibitemOpen
  \bibfield  {author} {\bibinfo {author} {\bibfnamefont {A.~B.}\ \bibnamefont
  {Balantekin}} \emph {et~al.},\ }\href@noop {} {\  (\bibinfo {year} {2022})},\
  \Eprint {http://arxiv.org/abs/2209.06872} {arXiv:2209.06872 [hep-ex]}
  \BibitemShut {NoStop}%
\bibitem [{\citenamefont {Rein}\ and\ \citenamefont
  {Sehgal}(1981)}]{bib:ReinSehgal}%
  \BibitemOpen
  \bibfield  {author} {\bibinfo {author} {\bibfnamefont {D.}~\bibnamefont
  {Rein}}\ and\ \bibinfo {author} {\bibfnamefont {L.~M.}\ \bibnamefont
  {Sehgal}},\ }\href {\doibase 10.1016/0003-4916(81)90242-6} {\bibfield
  {journal} {\bibinfo  {journal} {Ann. Phys.}\ }\textbf {\bibinfo {volume}
  {133}},\ \bibinfo {pages} {79} (\bibinfo {year} {1981})}\BibitemShut
  {NoStop}%
\bibitem [{\citenamefont {Leitner}\ \emph {et~al.}(2009)\citenamefont
  {Leitner}, \citenamefont {Buss}, \citenamefont {Alvarez-Ruso},\ and\
  \citenamefont {Mosel}}]{bib:REStheory}%
  \BibitemOpen
  \bibfield  {author} {\bibinfo {author} {\bibfnamefont {T.}~\bibnamefont
  {Leitner}}, \bibinfo {author} {\bibfnamefont {O.}~\bibnamefont {Buss}},
  \bibinfo {author} {\bibfnamefont {L.}~\bibnamefont {Alvarez-Ruso}}, \ and\
  \bibinfo {author} {\bibfnamefont {U.}~\bibnamefont {Mosel}},\ }\href
  {\doibase 10.1103/PhysRevC.79.034601} {\bibfield  {journal} {\bibinfo
  {journal} {Phys. Rev. C}\ }\textbf {\bibinfo {volume} {79}},\ \bibinfo
  {pages} {034601} (\bibinfo {year} {2009})},\ \Eprint
  {http://arxiv.org/abs/0812.0587} {arXiv:0812.0587 [nucl-th]} \BibitemShut
  {NoStop}%
\bibitem [{\citenamefont {Workman}\ \emph {et~al.}(2022)\citenamefont {Workman}
  \emph {et~al.}}]{bib:PDG}%
  \BibitemOpen
  \bibfield  {author} {\bibinfo {author} {\bibfnamefont {R.~L.}\ \bibnamefont
  {Workman}} \emph {et~al.} (\bibinfo {collaboration} {Particle Data Group}),\
  }\href {\doibase 10.1093/ptep/ptac097} {\bibfield  {journal} {\bibinfo
  {journal} {PTEP}\ }\textbf {\bibinfo {volume} {2022}},\ \bibinfo {pages}
  {083C01} (\bibinfo {year} {2022})}\BibitemShut {NoStop}%
\bibitem [{\citenamefont {Wittek}\ \emph {et~al.}(1989)\citenamefont {Wittek}
  \emph {et~al.}}]{bib:BEBCWA59}%
  \BibitemOpen
  \bibfield  {author} {\bibinfo {author} {\bibfnamefont {W.}~\bibnamefont
  {Wittek}} \emph {et~al.} (\bibinfo {collaboration} {BEBC WA59}),\ }\href
  {\doibase 10.1007/BF01557323} {\bibfield  {journal} {\bibinfo  {journal} {Z.
  Phys. C}\ }\textbf {\bibinfo {volume} {44}},\ \bibinfo {pages} {175}
  (\bibinfo {year} {1989})}\BibitemShut {NoStop}%
\bibitem [{\citenamefont {Kochanek}(2015)}]{bib:ICARUS_eta}%
  \BibitemOpen
  \bibfield  {author} {\bibinfo {author} {\bibfnamefont {I.}~\bibnamefont
  {Kochanek}},\ }\href@noop {} {\bibfield  {journal} {\bibinfo  {journal} {PhD,
  Silesia University, Katowice}\ } (\bibinfo {year} {2015})}\BibitemShut
  {NoStop}%
\bibitem [{\citenamefont {Rafi~Alam}\ \emph {et~al.}(2013)\citenamefont
  {Rafi~Alam}, \citenamefont {Sajjad~Athar}, \citenamefont {Alvarez-Ruso},
  \citenamefont {Ruiz~Simo}, \citenamefont {Vicente~Vacas},\ and\ \citenamefont
  {Singh}}]{bib:etatheory1}%
  \BibitemOpen
  \bibfield  {author} {\bibinfo {author} {\bibfnamefont {M.}~\bibnamefont
  {Rafi~Alam}}, \bibinfo {author} {\bibfnamefont {M.}~\bibnamefont
  {Sajjad~Athar}}, \bibinfo {author} {\bibfnamefont {L.}~\bibnamefont
  {Alvarez-Ruso}}, \bibinfo {author} {\bibfnamefont {I.}~\bibnamefont
  {Ruiz~Simo}}, \bibinfo {author} {\bibfnamefont {M.~J.}\ \bibnamefont
  {Vicente~Vacas}}, \ and\ \bibinfo {author} {\bibfnamefont {S.~K.}\
  \bibnamefont {Singh}},\ }in\ \href@noop {} {\emph {\bibinfo {booktitle}
  {{15th International Workshop on Neutrino Factories, Super Beams and Beta
  Beams}}}}\ (\bibinfo {year} {2013})\ \Eprint {http://arxiv.org/abs/1311.2293}
  {arXiv:1311.2293 [hep-ph]} \BibitemShut {NoStop}%
\bibitem [{\citenamefont {Fatima}\ \emph {et~al.}(2022)\citenamefont {Fatima},
  \citenamefont {Sajjad~Athar},\ and\ \citenamefont {Singh}}]{bib:etatheory2}%
  \BibitemOpen
  \bibfield  {author} {\bibinfo {author} {\bibfnamefont {A.}~\bibnamefont
  {Fatima}}, \bibinfo {author} {\bibfnamefont {M.}~\bibnamefont
  {Sajjad~Athar}}, \ and\ \bibinfo {author} {\bibfnamefont {S.~K.}\
  \bibnamefont {Singh}},\ }\href@noop {} {\  (\bibinfo {year} {2022})},\
  \Eprint {http://arxiv.org/abs/2211.08830} {arXiv:2211.08830 [hep-ph]}
  \BibitemShut {NoStop}%
\bibitem [{\citenamefont {Nakamura}\ \emph {et~al.}(2015)\citenamefont
  {Nakamura}, \citenamefont {Kamano},\ and\ \citenamefont
  {Sato}}]{bib:etatheory3}%
  \BibitemOpen
  \bibfield  {author} {\bibinfo {author} {\bibfnamefont {S.~X.}\ \bibnamefont
  {Nakamura}}, \bibinfo {author} {\bibfnamefont {H.}~\bibnamefont {Kamano}}, \
  and\ \bibinfo {author} {\bibfnamefont {T.}~\bibnamefont {Sato}},\ }\href
  {\doibase 10.1103/PhysRevD.92.074024} {\bibfield  {journal} {\bibinfo
  {journal} {Phys. Rev. D}\ }\textbf {\bibinfo {volume} {92}},\ \bibinfo
  {pages} {074024} (\bibinfo {year} {2015})},\ \Eprint
  {http://arxiv.org/abs/1506.03403} {arXiv:1506.03403 [hep-ph]} \BibitemShut
  {NoStop}%
\bibitem [{\citenamefont {Stancu}(2001)}]{bib:bnb}%
  \BibitemOpen
  \bibfield  {author} {\bibinfo {author} {\bibfnamefont {I.}~\bibnamefont
  {Stancu}},\ }\href {\doibase 10.2172/1212167} {\bibfield  {journal} {\bibinfo
   {journal} {FERMILAB-DESIGN-2001-03}\ } (\bibinfo {year} {2001}),\
  10.2172/1212167}\BibitemShut {NoStop}%
\bibitem [{\citenamefont {Abe}\ \emph {et~al.}(2017)\citenamefont {Abe} \emph
  {et~al.}}]{bib:superk_pdecay}%
  \BibitemOpen
  \bibfield  {author} {\bibinfo {author} {\bibfnamefont {K.}~\bibnamefont
  {Abe}} \emph {et~al.} (\bibinfo {collaboration} {Super-Kamiokande}),\ }\href
  {\doibase 10.1103/PhysRevD.96.012003} {\bibfield  {journal} {\bibinfo
  {journal} {Phys. Rev. D}\ }\textbf {\bibinfo {volume} {96}},\ \bibinfo
  {pages} {012003} (\bibinfo {year} {2017})},\ \Eprint
  {http://arxiv.org/abs/1705.07221} {arXiv:1705.07221 [hep-ex]} \BibitemShut
  {NoStop}%
\bibitem [{\citenamefont {Adams}\ \emph
  {et~al.}(2020{\natexlab{a}})\citenamefont {Adams} \emph
  {et~al.}}]{bib:uB_pi0}%
  \BibitemOpen
  \bibfield  {author} {\bibinfo {author} {\bibfnamefont {C.}~\bibnamefont
  {Adams}} \emph {et~al.} (\bibinfo {collaboration} {MicroBooNE}),\ }\href
  {\doibase 10.1088/1748-0221/15/02/P02007} {\bibfield  {journal} {\bibinfo
  {journal} {JINST}\ }\textbf {\bibinfo {volume} {15}},\ \bibinfo {pages}
  {P02007} (\bibinfo {year} {2020}{\natexlab{a}})},\ \Eprint
  {http://arxiv.org/abs/1910.02166} {arXiv:1910.02166 [hep-ex]} \BibitemShut
  {NoStop}%
\bibitem [{\citenamefont {Acciarri}\ \emph
  {et~al.}(2017{\natexlab{a}})\citenamefont {Acciarri} \emph
  {et~al.}}]{bib:uB_detector}%
  \BibitemOpen
  \bibfield  {author} {\bibinfo {author} {\bibfnamefont {R.}~\bibnamefont
  {Acciarri}} \emph {et~al.} (\bibinfo {collaboration} {MicroBooNE}),\ }\href
  {\doibase 10.1088/1748-0221/12/02/P02017} {\bibfield  {journal} {\bibinfo
  {journal} {JINST}\ }\textbf {\bibinfo {volume} {12}},\ \bibinfo {pages}
  {P02017} (\bibinfo {year} {2017}{\natexlab{a}})},\ \Eprint
  {http://arxiv.org/abs/1612.05824} {arXiv:1612.05824 [physics.ins-det]}
  \BibitemShut {NoStop}%
\bibitem [{\citenamefont {Snider}\ and\ \citenamefont
  {Petrillo}(2017)}]{bib:larsoft}%
  \BibitemOpen
  \bibfield  {author} {\bibinfo {author} {\bibfnamefont {E.~L.}\ \bibnamefont
  {Snider}}\ and\ \bibinfo {author} {\bibfnamefont {G.}~\bibnamefont
  {Petrillo}},\ }\href {\doibase 10.1088/1742-6596/898/4/042057} {\bibfield
  {journal} {\bibinfo  {journal} {J. Phys. Conf. Ser.}\ }\textbf {\bibinfo
  {volume} {898}},\ \bibinfo {pages} {042057} (\bibinfo {year}
  {2017})}\BibitemShut {NoStop}%
\bibitem [{\citenamefont {Aguilar-Arevalo}\ \emph {et~al.}(2009)\citenamefont
  {Aguilar-Arevalo} \emph {et~al.}}]{bib:mbflux}%
  \BibitemOpen
  \bibfield  {author} {\bibinfo {author} {\bibfnamefont {A.~A.}\ \bibnamefont
  {Aguilar-Arevalo}} \emph {et~al.} (\bibinfo {collaboration} {MiniBooNE}),\
  }\href {\doibase 10.1103/PhysRevD.79.072002} {\bibfield  {journal} {\bibinfo
  {journal} {Phys. Rev. D}\ }\textbf {\bibinfo {volume} {79}},\ \bibinfo
  {pages} {072002} (\bibinfo {year} {2009})}\BibitemShut {NoStop}%
\bibitem [{\citenamefont {Tena-Vidal}\ \emph {et~al.}(2021)\citenamefont
  {Tena-Vidal} \emph {et~al.}}]{bib:GENIEv3}%
  \BibitemOpen
  \bibfield  {author} {\bibinfo {author} {\bibfnamefont {J.}~\bibnamefont
  {Tena-Vidal}} \emph {et~al.} (\bibinfo {collaboration} {GENIE}),\ }\href
  {\doibase 10.1103/PhysRevD.104.072009} {\bibfield  {journal} {\bibinfo
  {journal} {Phys. Rev. D}\ }\textbf {\bibinfo {volume} {104}},\ \bibinfo
  {pages} {072009} (\bibinfo {year} {2021})},\ \Eprint
  {http://arxiv.org/abs/2104.09179} {arXiv:2104.09179 [hep-ph]} \BibitemShut
  {NoStop}%
\bibitem [{\citenamefont {Abe}\ \emph {et~al.}(2016)\citenamefont {Abe} \emph
  {et~al.}}]{bib:t2ktunedata}%
  \BibitemOpen
  \bibfield  {author} {\bibinfo {author} {\bibfnamefont {K.}~\bibnamefont
  {Abe}} \emph {et~al.} (\bibinfo {collaboration} {T2K}),\ }\href@noop {}
  {\bibfield  {journal} {\bibinfo  {journal} {Phys. Rev. D}\ }\textbf {\bibinfo
  {volume} {23}},\ \bibinfo {pages} {112012} (\bibinfo {year}
  {2016})}\BibitemShut {NoStop}%
\bibitem [{\citenamefont {Abratenko}\ \emph
  {et~al.}(2022{\natexlab{a}})\citenamefont {Abratenko} \emph
  {et~al.}}]{bib:uB_genietune}%
  \BibitemOpen
  \bibfield  {author} {\bibinfo {author} {\bibfnamefont {P.}~\bibnamefont
  {Abratenko}} \emph {et~al.} (\bibinfo {collaboration} {MicroBooNE}),\ }\href
  {\doibase 10.1103/PhysRevD.105.072001} {\bibfield  {journal} {\bibinfo
  {journal} {Phys. Rev. D}\ }\textbf {\bibinfo {volume} {105}},\ \bibinfo
  {pages} {072001} (\bibinfo {year} {2022}{\natexlab{a}})},\ \Eprint
  {http://arxiv.org/abs/2110.14028} {arXiv:2110.14028 [hep-ex]} \BibitemShut
  {NoStop}%
\bibitem [{\citenamefont {Agostinelli}\ \emph {et~al.}(2003)\citenamefont
  {Agostinelli} \emph {et~al.}}]{bib:geant4}%
  \BibitemOpen
  \bibfield  {author} {\bibinfo {author} {\bibfnamefont {S.}~\bibnamefont
  {Agostinelli}} \emph {et~al.} (\bibinfo {collaboration} {GEANT4}),\ }\href
  {\doibase 10.1016/S0168-9002(03)01368-8} {\bibfield  {journal} {\bibinfo
  {journal} {Nucl. Instrum. Meth. A}\ }\textbf {\bibinfo {volume} {506}},\
  \bibinfo {pages} {250} (\bibinfo {year} {2003})}\BibitemShut {NoStop}%
\bibitem [{\citenamefont {Acciarri}\ \emph
  {et~al.}(2017{\natexlab{b}})\citenamefont {Acciarri} \emph
  {et~al.}}]{bib:uB_noise}%
  \BibitemOpen
  \bibfield  {author} {\bibinfo {author} {\bibfnamefont {R.}~\bibnamefont
  {Acciarri}} \emph {et~al.} (\bibinfo {collaboration} {MicroBooNE}),\ }\href
  {\doibase 10.1088/1748-0221/12/08/P08003} {\bibfield  {journal} {\bibinfo
  {journal} {JINST}\ }\textbf {\bibinfo {volume} {12}},\ \bibinfo {pages}
  {P08003} (\bibinfo {year} {2017}{\natexlab{b}})},\ \Eprint
  {http://arxiv.org/abs/1705.07341} {arXiv:1705.07341 [physics.ins-det]}
  \BibitemShut {NoStop}%
\bibitem [{\citenamefont {Adams}\ \emph
  {et~al.}(2018{\natexlab{a}})\citenamefont {Adams} \emph
  {et~al.}}]{bib:uB_signal1}%
  \BibitemOpen
  \bibfield  {author} {\bibinfo {author} {\bibfnamefont {C.}~\bibnamefont
  {Adams}} \emph {et~al.} (\bibinfo {collaboration} {MicroBooNE}),\ }\href
  {\doibase 10.1088/1748-0221/13/07/P07006} {\bibfield  {journal} {\bibinfo
  {journal} {JINST}\ }\textbf {\bibinfo {volume} {13}},\ \bibinfo {pages}
  {P07006} (\bibinfo {year} {2018}{\natexlab{a}})},\ \Eprint
  {http://arxiv.org/abs/1802.08709} {arXiv:1802.08709 [physics.ins-det]}
  \BibitemShut {NoStop}%
\bibitem [{\citenamefont {Adams}\ \emph
  {et~al.}(2018{\natexlab{b}})\citenamefont {Adams} \emph
  {et~al.}}]{bib:uB_signal2}%
  \BibitemOpen
  \bibfield  {author} {\bibinfo {author} {\bibfnamefont {C.}~\bibnamefont
  {Adams}} \emph {et~al.} (\bibinfo {collaboration} {MicroBooNE}),\ }\href
  {\doibase 10.1088/1748-0221/13/07/P07007} {\bibfield  {journal} {\bibinfo
  {journal} {JINST}\ }\textbf {\bibinfo {volume} {13}},\ \bibinfo {pages}
  {P07007} (\bibinfo {year} {2018}{\natexlab{b}})},\ \Eprint
  {http://arxiv.org/abs/1804.02583} {arXiv:1804.02583 [physics.ins-det]}
  \BibitemShut {NoStop}%
\bibitem [{\citenamefont {Acciarri}\ \emph {et~al.}(2018)\citenamefont
  {Acciarri} \emph {et~al.}}]{bib:pandora}%
  \BibitemOpen
  \bibfield  {author} {\bibinfo {author} {\bibfnamefont {R.}~\bibnamefont
  {Acciarri}} \emph {et~al.} (\bibinfo {collaboration} {MicroBooNE}),\ }\href
  {\doibase 10.1140/epjc/s10052-017-5481-6} {\bibfield  {journal} {\bibinfo
  {journal} {Eur. Phys. J. C}\ }\textbf {\bibinfo {volume} {78}},\ \bibinfo
  {pages} {82} (\bibinfo {year} {2018})}\BibitemShut {NoStop}%
\bibitem [{\citenamefont {Adams}\ \emph
  {et~al.}(2020{\natexlab{b}})\citenamefont {Adams} \emph
  {et~al.}}]{bib:uB_SCE1}%
  \BibitemOpen
  \bibfield  {author} {\bibinfo {author} {\bibfnamefont {C.}~\bibnamefont
  {Adams}} \emph {et~al.} (\bibinfo {collaboration} {MicroBooNE}),\ }\href
  {\doibase 10.1088/1748-0221/15/07/P07010} {\bibfield  {journal} {\bibinfo
  {journal} {JINST}\ }\textbf {\bibinfo {volume} {15}},\ \bibinfo {pages}
  {P07010} (\bibinfo {year} {2020}{\natexlab{b}})},\ \Eprint
  {http://arxiv.org/abs/1910.01430} {arXiv:1910.01430 [physics.ins-det]}
  \BibitemShut {NoStop}%
\bibitem [{\citenamefont {Abratenko}\ \emph {et~al.}(2020)\citenamefont
  {Abratenko} \emph {et~al.}}]{bib:uB_SCE2}%
  \BibitemOpen
  \bibfield  {author} {\bibinfo {author} {\bibfnamefont {P.}~\bibnamefont
  {Abratenko}} \emph {et~al.} (\bibinfo {collaboration} {MicroBooNE}),\ }\href
  {\doibase 10.1088/1748-0221/15/12/P12037} {\bibfield  {journal} {\bibinfo
  {journal} {JINST}\ }\textbf {\bibinfo {volume} {15}},\ \bibinfo {pages}
  {P12037} (\bibinfo {year} {2020})},\ \Eprint
  {http://arxiv.org/abs/2008.09765} {arXiv:2008.09765 [physics.ins-det]}
  \BibitemShut {NoStop}%
\bibitem [{\citenamefont {Adams}\ \emph
  {et~al.}(2020{\natexlab{c}})\citenamefont {Adams} \emph
  {et~al.}}]{bib:uB_TPCcalib}%
  \BibitemOpen
  \bibfield  {author} {\bibinfo {author} {\bibfnamefont {C.}~\bibnamefont
  {Adams}} \emph {et~al.} (\bibinfo {collaboration} {MicroBooNE}),\ }\href
  {\doibase 10.1088/1748-0221/15/03/P03022} {\bibfield  {journal} {\bibinfo
  {journal} {JINST}\ }\textbf {\bibinfo {volume} {15}},\ \bibinfo {pages}
  {P03022} (\bibinfo {year} {2020}{\natexlab{c}})},\ \Eprint
  {http://arxiv.org/abs/1907.11736} {arXiv:1907.11736 [physics.ins-det]}
  \BibitemShut {NoStop}%
\bibitem [{\citenamefont {Abratenko}\ \emph
  {et~al.}(2022{\natexlab{b}})\citenamefont {Abratenko} \emph
  {et~al.}}]{bib:uB_PeLEE}%
  \BibitemOpen
  \bibfield  {author} {\bibinfo {author} {\bibfnamefont {P.}~\bibnamefont
  {Abratenko}} \emph {et~al.} (\bibinfo {collaboration} {MicroBooNE}),\ }\href
  {\doibase 10.1103/PhysRevD.105.112004} {\bibfield  {journal} {\bibinfo
  {journal} {Phys. Rev. D}\ }\textbf {\bibinfo {volume} {105}},\ \bibinfo
  {pages} {112004} (\bibinfo {year} {2022}{\natexlab{b}})},\ \Eprint
  {http://arxiv.org/abs/2110.14065} {arXiv:2110.14065 [hep-ex]} \BibitemShut
  {NoStop}%
\bibitem [{\citenamefont {Van De~Pontseele}(2020)}]{bib:Wouter}%
  \BibitemOpen
  \bibfield  {author} {\bibinfo {author} {\bibfnamefont {W.}~\bibnamefont {Van
  De~Pontseele}},\ }\emph {\bibinfo {title} {{Search for Electron Neutrino
  Anomalies with the MicroBooNE Detector}}},\ \href {\doibase 10.2172/1640226}
  {Ph.D. thesis},\ \bibinfo  {school} {Oxford U.} (\bibinfo {year}
  {2020})\BibitemShut {NoStop}%
\bibitem [{\citenamefont {Abratenko}\ \emph {et~al.}()\citenamefont {Abratenko}
  \emph {et~al.}}]{bib:supplementary}%
  \BibitemOpen
  \bibfield  {author} {\bibinfo {author} {\bibfnamefont {P.}~\bibnamefont
  {Abratenko}} \emph {et~al.} (\bibinfo {collaboration} {MicroBooNE}),\
  }\href@noop {} {\bibinfo  {journal} {Supplemental Material at [URL to be
  inserted by publisher]}\ }\BibitemShut {NoStop}%
\bibitem [{\citenamefont {Abratenko}\ \emph
  {et~al.}(2022{\natexlab{c}})\citenamefont {Abratenko} \emph
  {et~al.}}]{bib:uB_wiremod}%
  \BibitemOpen
\bibfield  {journal} {  }\bibfield  {author} {\bibinfo {author} {\bibfnamefont
  {P.}~\bibnamefont {Abratenko}} \emph {et~al.} (\bibinfo {collaboration}
  {MicroBooNE}),\ }\href {\doibase 10.1140/epjc/s10052-022-10270-8} {\bibfield
  {journal} {\bibinfo  {journal} {Eur. Phys. J. C}\ }\textbf {\bibinfo {volume}
  {82}},\ \bibinfo {pages} {454} (\bibinfo {year} {2022}{\natexlab{c}})},\
  \Eprint {http://arxiv.org/abs/2111.03556} {arXiv:2111.03556 [hep-ex]}
  \BibitemShut {NoStop}%
\bibitem [{\citenamefont {Golan}\ \emph {et~al.}(2012)\citenamefont {Golan},
  \citenamefont {Sobczyk},\ and\ \citenamefont {Zmuda}}]{bib:nuwro}%
  \BibitemOpen
  \bibfield  {author} {\bibinfo {author} {\bibfnamefont {T.}~\bibnamefont
  {Golan}}, \bibinfo {author} {\bibfnamefont {J.~T.}\ \bibnamefont {Sobczyk}},
  \ and\ \bibinfo {author} {\bibfnamefont {J.}~\bibnamefont {Zmuda}},\ }\href
  {\doibase 10.1016/j.nuclphysbps.2012.09.136} {\bibfield  {journal} {\bibinfo
  {journal} {Nucl. Phys. B Proc. Suppl.}\ }\textbf {\bibinfo {volume} {229}},\
  \bibinfo {pages} {499} (\bibinfo {year} {2012})}\BibitemShut {NoStop}%
\bibitem [{\citenamefont {Hayato}(2009)}]{bib:NEUT}%
  \BibitemOpen
  \bibfield  {author} {\bibinfo {author} {\bibfnamefont {Y.}~\bibnamefont
  {Hayato}},\ }\href@noop {} {\bibfield  {journal} {\bibinfo  {journal} {Acta
  Phys. Polon. B}\ }\textbf {\bibinfo {volume} {40}},\ \bibinfo {pages} {2477}
  (\bibinfo {year} {2009})}\BibitemShut {NoStop}%
\bibitem [{\citenamefont {Abratenko}\ \emph {et~al.}(2021)\citenamefont
  {Abratenko} \emph {et~al.}}]{bib:uB_PID}%
  \BibitemOpen
  \bibfield  {author} {\bibinfo {author} {\bibfnamefont {P.}~\bibnamefont
  {Abratenko}} \emph {et~al.} (\bibinfo {collaboration} {MicroBooNE}),\ }\href
  {\doibase 10.1007/JHEP12(2021)153} {\bibfield  {journal} {\bibinfo  {journal}
  {JHEP}\ }\textbf {\bibinfo {volume} {12}},\ \bibinfo {pages} {153} (\bibinfo
  {year} {2021})},\ \Eprint {http://arxiv.org/abs/2109.02460} {arXiv:2109.02460
  [physics.ins-det]} \BibitemShut {NoStop}%
\bibitem [{\citenamefont {Machado}\ \emph {et~al.}(2019)\citenamefont
  {Machado}, \citenamefont {Palamara},\ and\ \citenamefont
  {Schmitz}}]{bib:SBN_REVIEW}%
  \BibitemOpen
  \bibfield  {author} {\bibinfo {author} {\bibfnamefont {P.~A.}\ \bibnamefont
  {Machado}}, \bibinfo {author} {\bibfnamefont {O.}~\bibnamefont {Palamara}}, \
  and\ \bibinfo {author} {\bibfnamefont {D.~W.}\ \bibnamefont {Schmitz}},\
  }\href {\doibase 10.1146/annurev-nucl-101917-020949} {\bibfield  {journal}
  {\bibinfo  {journal} {Ann. Rev. Nucl. Part. Sci.}\ }\textbf {\bibinfo
  {volume} {69}},\ \bibinfo {pages} {363} (\bibinfo {year} {2019})},\ \Eprint
  {http://arxiv.org/abs/1903.04608} {arXiv:1903.04608 [hep-ex]} \BibitemShut
  {NoStop}%
\bibitem [{\citenamefont {Abed~Abud}\ \emph {et~al.}(2021)\citenamefont
  {Abed~Abud} \emph {et~al.}}]{bib:DUNE_ND}%
  \BibitemOpen
  \bibfield  {author} {\bibinfo {author} {\bibfnamefont {A.}~\bibnamefont
  {Abed~Abud}} \emph {et~al.} (\bibinfo {collaboration} {DUNE}),\ }\href
  {\doibase 10.3390/instruments5040031} {\bibfield  {journal} {\bibinfo
  {journal} {Instruments}\ }\textbf {\bibinfo {volume} {5}},\ \bibinfo {pages}
  {31} (\bibinfo {year} {2021})},\ \Eprint {http://arxiv.org/abs/2103.13910}
  {arXiv:2103.13910 [physics.ins-det]} \BibitemShut {NoStop}%
\end{thebibliography}%

\end{document}